\documentclass[12pt,a4]{article}
\usepackage{amsmath}
\usepackage{bm}
\usepackage{amsfonts}
\usepackage{amssymb}
\usepackage{graphics}
\usepackage[normal]{caption2}
\usepackage{subfigure}
\usepackage{rotating}
\usepackage{citesort}
\def\be{\begin{equation}}
\def\ee{\end{equation}}
\def\ba{\begin{array}}
\def\ea{\end{array}}
\def\bea{\begin{eqnarray}}
\def\eea{\end{eqnarray}}

\begin{document}
\baselineskip 20pt \setlength\tabcolsep{2.5mm}
\renewcommand\arraystretch{1.5}
\setlength{\abovecaptionskip}{0.1cm}
\setlength{\belowcaptionskip}{0.5cm}
\begin{center} {\large\bf Study of participant-spectator matter, thermalization and other related phenomena for neutron-rich colliding pair}\\
\vspace*{0.4cm}
{\bf Sakshi Gautam} \footnote{Email:~sakshigautm@gmail.com}\\
{\it  Department of Physics, Panjab University, Chandigarh -160
014, India.\\}
\end{center}
We study the participant-spectator matter, density and temperature
reached in heavy-ion reactions of neutron-rich systems having N/Z
ratio varying from 1.0 to 2.0 at 50 and 250 MeV/nucleon. Our
results show a weak dependence  of these quantities on the neutron
content of the colliding systems. We also shed light on the role
of neutron content on the thermalization achieved in a reaction.


\newpage
\baselineskip 20pt
\section{Introduction}
\label{intro}
 Heavy-ion collisions (HIC) provide a unique
opportunity to study the properties of hot and dense nuclear
matter as the nuclear matter is produced at temperature, densities
and neutron-proton ratios away from that of the ground state
nuclei. The nuclear equation of state (EOS) as well as the
in-medium nucleon-nucleon (nn) cross section plays a great role in
determining these properties \cite{sch,doss86}. The importance of
collective transverse in-plane flow in determining these
properties has been known for past three decades. The collective
transverse in-plane flow has been found to be influenced by the
entrance channel parameters like mass of the colliding system
\cite{ogli89,andro03}, impact parameter
\cite{pan93,luka05,zhang06} and incident energy
\cite{zhang06,wmz90,luka08}. The competition between attractive
interactions due to mean field (at low incident energies) and
repulsive interactions due to nucleon-nucleon (nn) collisions (at
high incident energies) led to the disappearance of flow at a
particular incident energy (where attractive and repulsive
interactions counterbalance each other), known as energy of
vanishing flow (EVF) \cite{krof}. Since one of the motivations
behind studying HIC is to extract information about the hot and
dense nuclear matter, so few studies also shed light on the
properties such as density and temperature reached in HIC and also
on the participant-spectator matter \cite{sood2,sood3}. Puri and
collaborators have shown that participant-spectator matter can act
as a barometer for the study of EVF. The study also pointed
towards the insensitivity of participant-spectator matter towards
equation of state, momentum-dependent interactions (MDI) and nn
cross sections. These studies were done at the EVF. Moreover,
these studies were restricted to the systems lying close to the
stability line.
 \par With the existing and upcoming radioactive
ion beam (RIB) facilities around the world, the present physics
interest have been shifted towards nuclei far from the stability
line. Various phenomena like fusion-fission, collective flow, EVF
and multifragmentation have been studied for such nuclei which lie
far from the stability line. So, in the present paper, we aim to
see the properties like, density, temperature, thermalization, and
participant-spectator matter for neutron-rich systems lying far
from the stability line.
\par
The present study is carried out within the framework of
isospin-dependent quantum molecular dynamics (IQMD) model which is
described in detail in section 2. Our results along with
discussions are presented in section 3 and we summarized the
results in section4.


 \section{The model}
 \label{sec:1}
 The IQMD model \cite{hart98} which is the extension of quantum molecular dynamics (QMD) \cite{aichqmd}
 model treats different charge states of
nucleons, deltas, and pions explicitly, as inherited from the
Vlasov-Uehling-Uhlenbeck (VUU) model. The IQMD model has been used
successfully for the analysis of a large number of observables
from low to relativistic energies. Puri and coworkers have
demonstrated that QMD and IQMD carries essential physics needed to
demonstrate the various phenomena such as collective flow,
multifragmentation and particle production \cite{dhawan,batko}.
The isospin degree of freedom enters into the calculations via
symmetry potential, cross sections and Coulomb interactions.
 \par
 In this model, baryons are represented by Gaussian-shaped density distributions

\begin{equation}
f_{i}(\vec{r},\vec{p},t) =
\frac{1}{\pi^{2}\hbar^{2}}\exp(-[\vec{r}-\vec{r_{i}}(t)]^{2}\frac{1}{2L})
\times \exp(-[\vec{p}- \vec{p_{i}}(t)]^{2}\frac{2L}{\hbar^{2}})
 \end{equation}

 where L is the Gaussian width which can be regarded as
 a description of the interaction range of a particle.
 In IQMD the Gaussian width is system dependent.
 The system dependence of L in IQMD has been introduced in
 order to obtain the maximum stability of the nucleonic
 density distribution. for Au+Au a value of L = 8.66 fm$_{2}$ is chosen, for Ca+Ca and lighter nuclei L = 4.33 fm$_{2}$.
 Nucleons are initialized in a sphere with radius R = 1.12
A$^{1/3}$ fm, in accordance with liquid-drop model.
 Each nucleon occupies a volume of \emph{h$^{3}$}, so that phase space is uniformly filled.
 The initial momenta are randomly chosen between 0 and Fermi momentum ($\vec{p}$$_{F}$).
 The nucleons of the target and projectile interact by two- and three-body\textrm{ Skyrme} forces,
 \textrm{Yukawa} potential and \textrm{Coulomb} interactions and momentum-dependent interactions.
 In addition to the use of explicit charge states of all baryons
and mesons, a symmetry potential between protons and neutrons
 corresponding to the Bethe-Weizs\"acker mass formula has been included. The hadrons propagate using Hamilton equations of motion:

\begin {eqnarray}
\frac{d\vec{{r_{i}}}}{dt} = \frac{d\langle H
\rangle}{d\vec{p_{i}}};& & \frac{d\vec{p_{i}}}{dt} = -
\frac{d\langle H \rangle}{d\vec{r_{i}}}
\end {eqnarray}

 with

\begin {eqnarray}
\langle H\rangle& =&\langle T\rangle+\langle V \rangle
\nonumber\\
& =& \sum_{i}\frac{p^{2}_{i}}{2m_{i}} + \sum_{i}\sum_{j>i}\int
f_{i}(\vec{r},\vec{p},t)V^{\textrm{ij}}(\vec{r}~',\vec{r})
 \nonumber\\
& & \times f_{j}(\vec{r}~',\vec{p}~',t) d\vec{r}~ d\vec{r}~'~
d\vec{p}~ d\vec{p}~'.
\end {eqnarray}

 The baryon potential V$^{\textrm{ij}}$ in the above relation, reads as

 \begin {eqnarray}
  \nonumber V^{ij}(\vec{r}~'-\vec{r})& =&V^{ij}_{Skyrme} + V^{ij}_{Yukawa} +
  V^{ij}_{Coul} + V^{ij}_{mdi} + V^{ij}_{sym}
    \nonumber\\
   & =& [t_{1}\delta(\vec{r}~'-\vec{r})+t_{2}\delta(\vec{r}~'-\vec{r})\rho^{\gamma-1}(\frac{\vec{r}~'+\vec{r}}{2})]
   \nonumber\\
   &  & +t_{3}\frac{\exp(|(\vec{r}~'-\vec{r})|/\mu)}{(|(\vec{r}~'-\vec{r})|/\mu)}+
    \frac{Z_{i}Z_{j}e^{2}}{|(\vec{r}~'-\vec{r})|}
   \nonumber \\
   &  & +t_{4}\ln^{2}[t_{5}(\vec{p}~'-\vec{p})^{2} +
    1]\delta(\vec{r}~'-\vec{r})
    \nonumber\\
   &  & +t_{6}\frac{1}{\varrho_{0}}T_{3i}T_{3j}\delta(\vec{r_{i}}~'-\vec{r_{j}}).
 \end {eqnarray}
The Skyrme potential V$^{ij}_{Skyrme}$ consists of two-body and
three-body interactions and can be parameterized as:
\begin {eqnarray}
U=\alpha \left({\frac {\rho}{\rho_{0}}}\right) + \beta
\left({\frac {\rho}{\rho_{0}}}\right)^{\gamma}
 \end {eqnarray}
 The parameters $\alpha$ and $\beta$ can be fixed by requirement
 that at normal nuclear matter density the average binding energy
 should be -16 MeV and total energy should have a minimum at
 $\rho_{0}$. Different values of $\gamma$ lead to different
 equation of states. Naturally, larger value of $\gamma$ leads to
 hard EOS whereas smaller value of $\gamma$ results in soft
 equation of state. The compressibility K (MeV) is 200 and 380 for
 soft and hard EOS, respectively.
Here t$_{6}$ = 100 MeV and \emph{Z$_{i}$} and \emph{Z$_{j}$}
denote the charges of \emph{ith} and \emph{jth} baryon, and
\emph{T$_{3i}$} and \emph{T$_{3j}$} are their respective
\emph{T$_{3}$} components (i.e., $1/2$ for protons and $-1/2$ for
neutrons). The parameters\emph{ $\mu$} and
\emph{t$_{1}$,....,t$_{6}$} are adjusted to the real part of the
nucleonic optical potential. The Yukawa potential
$V^{ij}_{Yukawa}$ in IQMD is very short ranged ($\mu$ = 0.4 fm)
and weak.
 For the density dependence of  the nucleon optical potential, standard \textrm{Skyrme} type parametrization is employed.
We use a soft equation of state with and without
momentum-dependent interactions (MDI) labeled as Soft and SMD,
respectively along with the standard isospin- and energy-dependent
nn cross section reduced by
  20$\%$, i.e. $\sigma$ = 0.8 $\sigma_{nn}^{free}$. The cross section for
  proton-neutron collisions is three times as that for proton-proton
  or neutron-neutron collisions. The cross sections for neutron-neutron
collisions are assumed to be equal to the proton-proton cross
sections. The more details about the elastic and inelastic cross
sections for proton-proton and proton-neutron collisions can be
found in \cite{hart98,cug}.  In a recent study, Gautam \emph{et
al}. \cite{gaum1}
  has confronted the theoretical calculations of IQMD with the data of $^{58}Ni+^{58}Ni$ and $^{58}Fe+^{58}Fe$ \cite{pak97}.
  The results with the soft EOS (along with the
momentum-dependent interactions) and above choice of cross section
are in good agreement with the data at all colliding geometries.
Two particles collide if their minimum distance\emph{ d} fulfills
\begin {equation}
 d \leq d_{0} = \sqrt{\frac{\sigma_{tot}}{\pi}},   \sigma_{tot} =
 \sigma(\sqrt{s}, type),
\end {equation}
where 'type' denotes the ingoing collision partners (N-N....).
Explicit Pauli blocking is also included; i.e. Pauli blocking of
the neutrons and protons is treated separately. We assume that
each nucleon occupies a sphere in coordinate and momentum space.
This trick yields the same Pauli blocking ratio as an exact
calculation of the overlap of the Gaussians will yield. We
calculate the fractions P$_{1}$ and P$_{2}$ of final phase space
for each of the two scattering partners that are already occupied
by other nucleons with the same isospin as that of scattered ones.
The collision is blocked with the probability
\begin {equation}
 P_{block} = 1-[1 - min(P_{1},1)][1 - min(P_{2},1)],
\end {equation}
and, correspondingly is allowed with the probability 1 -
P$_{block}$. For a nucleus in its ground state, we obtain an
averaged blocking probability $\langle P_{block}\rangle$ = 0.96.
Whenever an attempted collision is blocked, the scattering
partners maintain the original momenta prior to scattering.
\par
 \section{Results and discussion}
 \label{sec:2}

 We simulate the reactions of Ca+Ca and Xe+Xe series having
 N/Z ratio varying from 1.0 to 2.0 in steps of 0.2. In particular we simulate the
 reactions of $^{40}$Ca+$^{40}$Ca, $^{44}$Ca+$^{44}$Ca,
 $^{48}$Ca+$^{48}$Ca, $^{52}$Ca+$^{52}$Ca, $^{56}$Ca+$^{56}$Ca
 and \\ $^{60}$Ca+$^{60}$Ca and $^{110}$Xe+$^{110}$Xe,
 $^{120}$Xe+$^{120}$Xe, $^{129}$Xe+$^{129}$Xe,
 $^{140}$Xe+$^{140}$Xe, $^{151}$Xe+$^{151}$Xe, and
 $^{162}$Xe+$^{162}$Xe at b/b$_{max}$ = 0.2-0.4. The incident
 energies are taken to be 50 and 250 MeV/nucleon.
 \par
  In the present work, we define the participant and spectator matter
in terms of the nucleonic concept. All the nucleons which have
experienced at least one collision are labeled as
\emph{participant matter}, scaled to the total mass of the
reacting nuclei. The remaining matter is labeled as
\emph{spectator matter}. These definitions provide a possibility
to study the reaction in terms of participant-spectator fireball
model.

 \begin{figure}[!t]\centering
 \vskip 0.5cm
  \includegraphics[angle=0,width=10cm]{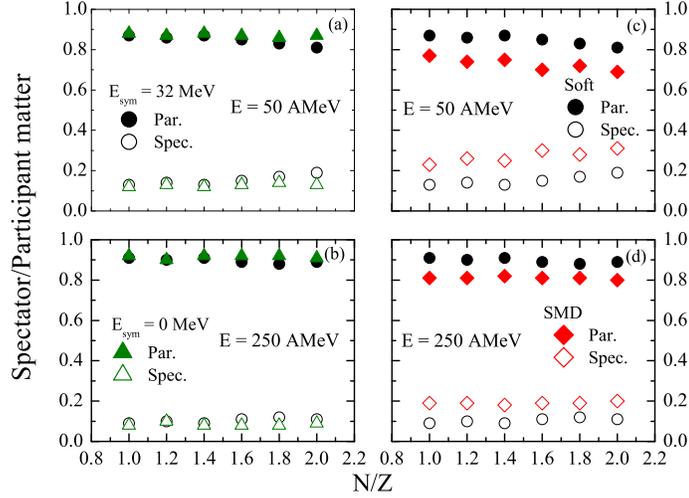}
 \caption{(Color online) The N/Z dependence of participant and
spectator matter for the reactions of Ca+Ca at 50 (upper panels)
and 250 (lower) MeV/nucleon for different N/Z ratios with
E$_{sym}$ = 0 MeV (left panels) and MDI (right). Various symbols
are explained in the text.}\label{fig1}
\end{figure}


\par
In fig. 1, we display the N/Z dependence of participant-spectator
matter for Ca+Ca reactions. Upper (lower) panels represent the
results for 50 (250) MeV/nucleon. Solid (open) symbols are for
participant (spectator) matter. From figure (circles), we find
that at lower incident energy there is a slight change in
participant-spectator matter with N/Z ratio whereas, at higher
incident energy of 250 MeV/ \\
nucleon, the participant-spectator matter is almost independent of
the neutron content of the colliding systems. As we are increasing
the neutron content at fixed Z, the mass of the system also
increases. Due to increase in mass, the participant matter should
increase with N/Z ratio (at fixed incident energy). But as we are
moving to higher neutron content, the role of symmetry energy also
increases, which being repulsive in nature leads to lesser number
of collisions in systems with more neutrons and hence participant
matter should decrease with increase in neutron content of the
system. So the net effect is due to the interplay of mass effects
and effect of the symmetry energy. To check the relative
importance of these two effects we make the strength of potential
part of symmetry energy zero and calculate the
participant-spectator matter for Ca+Ca reactions. The results are
displayed in Fig. 1(a) (50 MeV/nucleon) and 1(b) (250 MeV/nucleon)
(triangles). From figure, we see that on reducing the strength of
symmetry energy to zero, the participant (spectator) matter
increases (decreases) slightly for higher neutron content whereas
the effect on lower N/Z ratios is almost negligible.

\par
The momentum-dependent interactions also play a significant role
in the dynamics of heavy-ion collisions. To see the role of MDI on
participant-spectator matter we calculate the
participant-spectator matter for Soft+MDI (SMD) EOS for the
reactions of Ca+Ca. The results are displayed by diamonds in Fig.
1(c) and 1(d) at 50 and 250 MeV/nucleon, respectively. From
figure, we see that participant [closed diamonds] (spectator, open
diamonds) matter decreases (increases) with SMD. This is due to
the fact that since MDI is repulsive in nature, so the matter is
thrown away from the central dense zone and hence lesser number of
collisions will take place which leads to decrease (increase) in
participant (spectator) matter. Similar behaviour is observed for
Xe+Xe reactions (results not shown here).
  \begin{figure}[!t]\centering
 \vskip 0.5cm
  \includegraphics[angle=0,width=10cm]{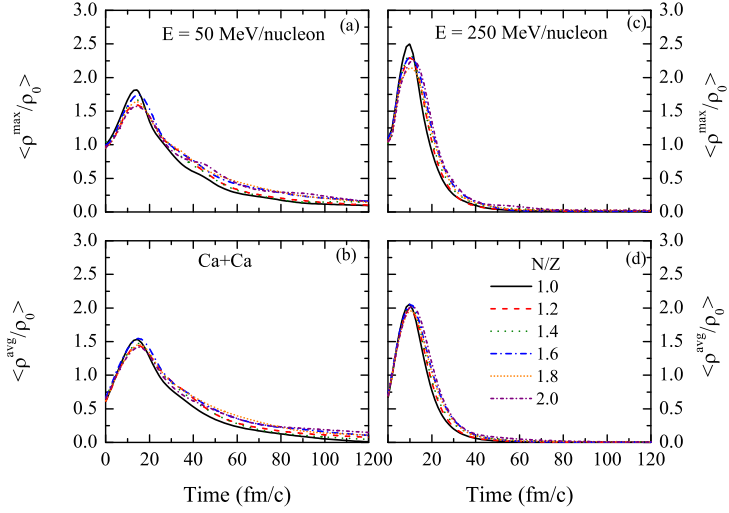}
 \caption{(Color online) The time evolution of the maximum
($<\rho_{max}/\rho_{0}>$) and average ($<\rho_{mavg}/\rho_{0}>$)
densities for the reactions of Ca+Ca at 50 (left panels) and 250
(right) MeV/nucleon. Lines have been explained in
text.}\label{fig2}
\end{figure}

\par
Since one of the motivations behind studying heavy-ion collisions
is to extract information about the hot and dense nuclear matter.
In our approach,  matter density is calculated by \cite {khoa921}
\begin{equation}
\rho(\vec{r},t)=\sum_{i=1}^{A_{T}+A_{P}}\frac{1}{(2\pi L)^{3/2}}
e^{(-(\vec{r}-\vec{r}_{i}(t))^{2}/2L)}.
\end{equation}
Here $A_{T}$ and $A_{P}$ stand, respectively, for the mass of the
target and projectile. In actual calculations, we take a sphere of
2 fm radius around the center-of-mass and compute the density at
each time step during the reaction using Eq. (5.1). Naturally, one
can either extract an average density $\langle \rho^{avg} \rangle$
over the whole sphere or a maximal value of the density $\langle
\rho^{max} \rangle$ reached anywhere in the sphere.
\par
In fig. 2, we display the time evolution of maximum (upper panels)
and average density (lower panels) at 50 (left panels) and 250
MeV/nucleon (right) for the reactionsof Ca+Ca. Solid, dashed,
dotted, dash-dotted, short dotted and short dash-dotted represent
N/Z ratios of 1.0, 1.2, 1.4, 1.6, 1.8, and 2.0, respectively. From
fig. 2 (left panels), we find that maximum and average density
rises as the reaction proceeds, reaches maximum at 10-20 fm/c when
the matter is highly compressed and finally decreases during the
expansion phase. The average density reached in the reactions is
about 1.5 times the normal nuclear matter density. The maximal
value of the density decreases with increase in neutron content of
the system. This slight decrease of the maximal value may be due
to the larger effect of symmetry energy in systems with higher
neutron content, which in turn, pushes the matter away from the
central dense zone due to the repulsive nature of symmetry energy.
From fig. 2 (right panels), we see that at 250 MeV/nucleon, much
higher density is achieved. This is because of the much violent
nature of the reaction at higher incident energy.
 \begin{figure}[!t]\centering
 \vskip 0.5cm
  \includegraphics[angle=0,width=10cm]{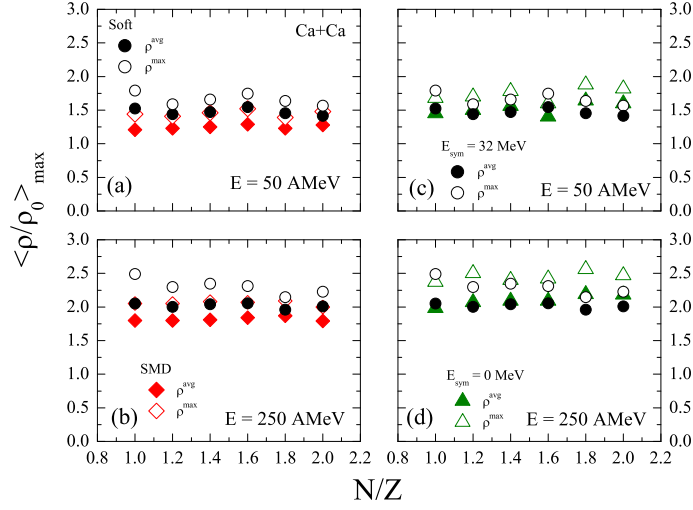}
 \caption{ (Color online) The N/Z dependence of maximal value of
maximum ($<\rho_{max}/\rho_{0}>$) and
 average density ($<\rho_{avg}/\rho_{0}>$) for the reactions of Ca+Ca at 50 (upper panels) and 250 (lower)
 MeV/nucleon with MDI (left panels) and E$_{sym}$ = 0 MeV (right). Symbols have same meaning as in Fig. 1.}\label{fig3}
\end{figure}

 \begin{figure}[!t]\centering
 \vskip 0.5cm
  \includegraphics[angle=0,width=10cm]{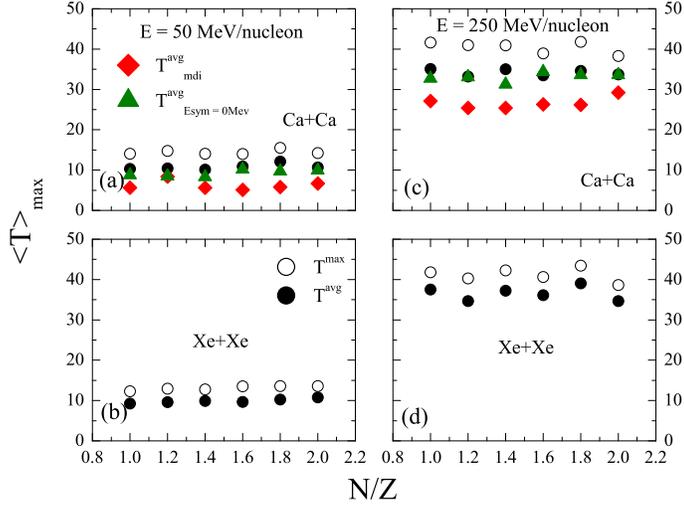}
 \caption{(Color online) The N/Z dependence of maximal value of
maximum (T$_{max}$)and average temperature (T$_{avg}$) for Ca+Ca
(upper panels) and Xe+Xe (lower) at 50 (left) and 250 (right)
MeV/nucleon. Diamonds and triangles represent the results for MDI
and E$_{sym}$ = 0 MeV, respectively.}
 \label{fig4}
\end{figure}

\par
In fig. 3, we display the N/Z dependence of the maximal value of
the maximum and average density achieved in a reaction at 50
(upper panels) and 250 (lower) MeV/nucleon for the reactions of
Ca+Ca reactions. Solid (open) circles represent the maximal value
of $\rho_{avg}$ ($\rho_{max}$). We find that maximal value almost
remains independent of the neutron content of the reacting system.
\par
To see the role of MDI and symmetry energy, we display the results
with MDI (Figs. 3(a) and 3(b)) (diamonds) and with no symmetry
energy (Figs. 3(c) and 3(d)) (triangles)  on average (closed
symbols) and maximum density (open symbols) achieved in reactions
of Ca+\\Ca. From figure, we see that average and maximum density
decreases with SMD as compared to soft EOS, although the N/Z
behavior remains independent. Similarly, when we reduce the
strength of symmetry energy to zero, the average and maximum
density increases slightly for higher neutron content whereas for
lower neutron content, the effect of symmetry energy is
negligible.

\par
The associated quantity linked with the dense matter is the
temperature. In principle, a true temperature can be defined only
for a thermalized and equilibrated matter. Since in heavy-ion
collisions the matter is non-equilibrated, one can not talk of
``temperature''. One can, however, look in terms of the local
environment only. In our present case, we follow the description
of the temperature given in Refs. \cite{khoa921,khoa922}.
 In the present case, extraction of the temperature {\it T} is based on the local
density approximation, i.e., one deduces the temperature in a
volume element surrounding the position of each particle at a
given time step \cite{khoa921,khoa922}. Here, we postulate that
each local volume element of nuclear matter in coordinate space
and time has some ``temperature'' defined by the diffused edge of
the deformed Fermi distribution consisting of two colliding Fermi
spheres, which is typical for a nonequilibrium momentum
distribution in heavy-ion collisions.

In this formalism (dubbed the hot Thomas-Fermi approach
\cite{khoa921}), one determines extensive quantities like the
density and kinetic energy as well as entropy with the help of
momentum distributions at a given temperature. Using this
formalism, we also extracted the average and maximum temperature
within a central sphere of 2 fm
radius as described in the case of density.\\
In fig. 4, we display the N/Z dependence of the temperature
achieved in a reaction. Upper (lower) panels display the results
for Ca+Ca (Xe+Xe) reactions. Solid (open) symbols represent the
results of T$^{avg}$ (T$^{max}$). Left (right) panels represent
the results of 50 (250) MeV/nucleon. We find that higher
temperatures are achieved in reactions taking place at higher
incident energy. For low incident energy, temperature remains
almost constant for lower neutron content and shows slight
increase for higher neutron content. We also study the effect of
MDI (diamonds) and symmetry energy (triangles) on average
temperature for the reactions of Ca+Ca at 50 and 250 MeV/nucleon.
From figure, we see that average temperature decreases for all N/Z
ratios when MDI is included. Similarly, when we reduce the
strength of symmetry energy to zero, there is a very slight
decrease in average temperature. The N/Z behavior, although,
remains unaltered. Similar behavior of MDI and symmetry energy is
observed for Xe+Xe reactions (results not shown here).

 \begin{figure}[!t]\centering
 \vskip 0.5cm
  \includegraphics[angle=0,width=10cm]{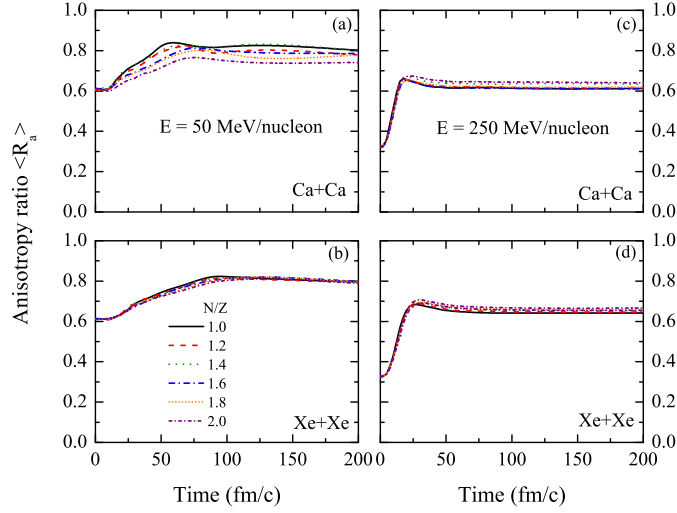}
 \caption{(Color online) The time evolution of anisotropy ratio
for Ca+Ca (upper panels) and Xe+Xe (lower) reactions at 50 (left
panels) and 250 (right) MeV/nucleon for different N/Z ratios.
Lines have same meaning as in Fig. 2.}\label{fig5}
\end{figure}
\par
In fig. 5 we display the time evolution of anisotropy ratio
$<R_{a}>$ reached in a reaction. The $<R_{a}>$ is defined as
\begin{equation}
<R_{a}> =
\frac{\sqrt{p_{x}^{2}}+\sqrt{p_{y}^{2}}}{2\sqrt{p_{z}^{2}}}.
\end{equation}

This anisotropy ratio is an indicator of the global equilibrium of
the system. This represents the equilibrium of the whole system
and does not depend on the local positions. The full global
equilibrium averaged over large number of events will correspond
to $<R_{a}>$ = 1. From figure, we see that as the reaction
proceeds, the $<R_{a}>$ ratio increases and saturates after the
high dense phase is over. We also find that the effect of N/Z
ratio is very less on the thermalization achieved in a reaction.
We also find a smaller value of $<R_{a}>$ at the start of the
reaction for higher energy projectile (see right panels).

\par
In fig. 6, we display the time evolution of relative momentum. The
 relative momentum $<K_{R}>$ of two
colliding Fermi spheres, is defined as

\begin{equation}
<K_{R}> = <|\vec{P}_{P}(\vec{r},t)-\vec{P}_{T}(\vec{r},t)|/\hbar>,
 \label{kr}
\end{equation}
 \begin{figure}[!t]\centering
 \vskip 0.5cm
  \includegraphics[angle=0,width=10cm]{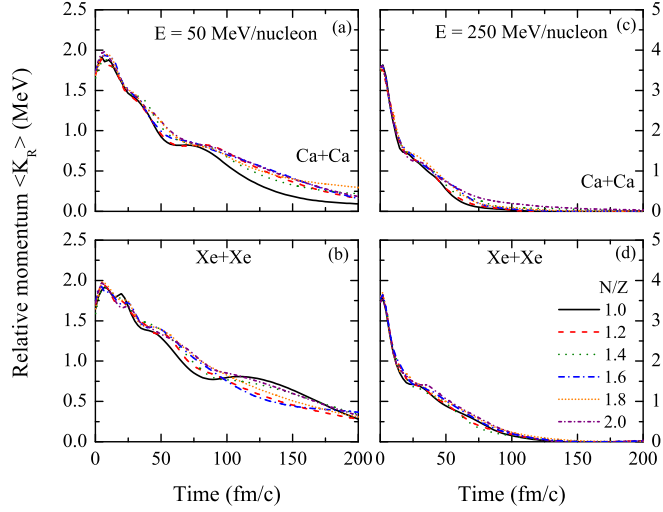}
\caption{(Color online) Same as Fig. 5, but for the time evolution
of relative momentum.}\label{fig6}
\end{figure}


where

\begin{equation}
\vec{P_{i}}(\vec{r},t) =
\frac{\sum_{j=1}^{A}\vec{P_{j}}(t)\rho_{j}(\vec{r},t)}
{\rho_{j}(\vec{r},t)}~~~~~         i=1,2,
\end{equation}
i.e., \emph{i} = 1 and 2 correspond to \emph{i} = P (Projectile)
and T(target). Here $\vec{P_{j}}$ and $\rho_{j}$ are the momentum
and density of the \emph{j}th particle and \emph{i} stands for
either projectile or target. The $<$$K_{R}$$>$ is an indicator of
the local equilibrium because it depends also on the local
position \emph{r}.

We find that relative momentum decreases as the reaction proceeds.
The effect of neutron content is very less on the value of
relative momentum achieved in a reaction. The effect diminishes
further at higher incident energy. Also, at higher incident
energy, at the start of the reaction we have a higher value of the
relative momentum. Also the decrease in $<$$K_{R}$$>$ is much
faster for higher incident energy.
\par
\section{Summary}
\label{sec:4}
 We studied the dependence of participant-spectator
 matter, temperature and density reached in heavy-ion reactions of
 neutron-rich systems on the neutron content of the colliding system. The study is done at 50 and 250
 MeV/nculeon. Our calculations showed that the behaviour of these
 quantities is almost independent of the neutron content of the system
 under study. In addition, we also studied the thermalization
 achieved in neutron-rich systems. Thermalization also showed
 insensitivity towards the neutron content of the reacting
 systems.
 \par
This work has been supported by a grant from Centre of Scientific
and Industrial Research (CSIR), Govt. of India. Author is thankful
to Profs. J. Aichelin and R. K. Puri for enlightening discussions
on the present work.

\end{document}